\documentclass[twocolumn,showpacs,superscriptaddress,prl]{revtex4-1}

\usepackage{graphicx}
\usepackage{dcolumn}
\usepackage{bm}
\usepackage{natbib}
\usepackage{subfigure}
\usepackage{amssymb,amsmath}
\usepackage{epstopdf}

\begin{document}


\title{Frustrated order on extrinsic geometries}%

\author{Badel L. Mbanga}
\email{badel@polysci.umass.edu}
\affiliation{Department of Polymer Science and Engineering, University of Massachusetts, Amherst, MA 01003.}
\affiliation{Department of Physics, University of Massachusetts, Amherst,
MA 01003}

\author{Gregory M. Grason}
 \email{grason@pse.umass.edu}
\affiliation{Department of Polymer Science and Engineering, University of Massachusetts, Amherst, MA 01003.}

\author{Christian D. Santangelo }
\email{csantang@physics.umass.edu}
\affiliation{Department of Physics, University of Massachusetts, Amherst,
MA 01003}
 \email{csantang@physics.umass.edu}

\begin{abstract}
We study, analytically and theoretically, defects in a nematically-ordered surface that couple to the extrinsic  geometry of a surface. Though the intrinsic geometry tends to confine topological defects to regions of large Gaussian curvature, extrinsic couplings tend to orient the nematic in the local direction of maximum or minimum bending. This additional frustration is unavoidable and most important on surfaces of negative Gaussian curvature, where it leads to a complex ground state thermodynamics. We show, in contradistinction to the well-known effects of intrinsic geometry, that extrinsic curvature expels disclinations from the region of maximum curvature above a critical coupling threshold. On catenoids lacking an ``inside-outside'' symmetry, defects are expelled altogether.

\end{abstract}

\date{\today}
\pacs{61.30.Dk, 61.30.Jf,61.30.Hn}

\maketitle

The intrinsic curvature of a surface frustrates the order of materials living upon it, making defects necessary even in the ground state~\cite{vitelli_turner}. This frustration derives from the incompatibility of straight and parallel directions on surfaces with Gaussian curvature.   A nematic texture on a sphere, for example, must have a net topological charge of $+2$~ \cite{prost_lubensky, nelson_colloids, shin}, as do, for example, lines of latitude on the globe.  On more complex surfaces with non-uniform intrinsic curvature, the ground-state organization of defects is well-described theoretically by competition between defect-defect interactions, which favor separation of like-charged defects, and curvature-defect interactions, which favor the localization of defects in regions of relatively high Gaussian curvature~\cite{nelson_peliti, vitelli_nelson}.  This picture is far from complete, however, as real surfaces are endowed with an extrinsic geometry that depends in detail on how they sit in space \cite{helfrich_prost, terentjev, huber}. Stripes on a curved surface, for example, tend to orient along``flat'' directions to minimize their bending energy, an effect observed in block copolymer films \cite{santangelo}.

\begin{figure}[b]
\begin{center}
\includegraphics[width=3.25in]{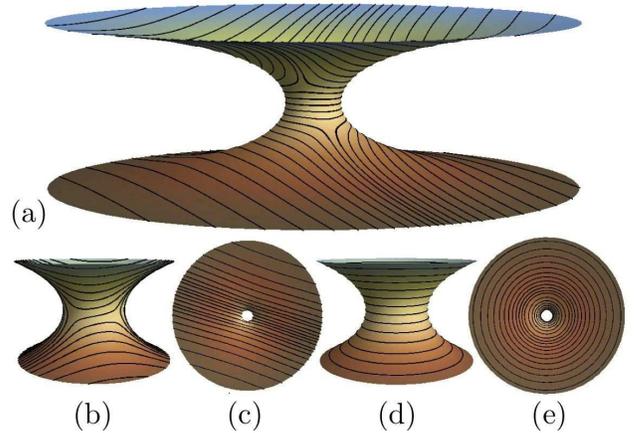} 
\caption{\label{fig:catenoid} (color online) (a) A catenoid with four, {\it partially expelled} $-1/2$ defects as determined by numerical minimization. Defects are confined to the neck in the ground state of intrinsic energy, shown in side- and top-view in (b) and (c).  In the ground state of extrinsic energy, (d) and (e), defects are expelled from the necked results in an overall $+1$ charge on the planar regions.} 
 \label{illustration}
\end{center}
\end{figure}

In this Letter, we demonstrate that the extrinsic geometry of a surface generates a strong and previously unstudied source of frustration which controls the ground state of anisotropically-ordered surfaces. While the intimate relationship between topological defects and intrinsic curvature is well-established, we show that the coupling between the spatial organization of defects and extrinsic geometry is as important, and in some cases {\it more important}, than intrinsic geometry for determining the ground state of nematic order on anisotropic surfaces.  Away from umbilics, extrinsic geometry gives rise to symmetry-breaking fields in regions of high curvature that favor uniform order most strongly in precisely those regions where the intrinsic geometry favors the locally ``isotropic" textures generated by defects.  The frustration between textures favorable to intrinsic and extrinsic energies becomes particularly pronounced on surfaces of negative Gaussian curvature.  As prototypical example this frustration, we study the ground state of nematic order on the surface of catenoid (see Fig. \ref{fig:catenoid}).  While the intrinsic curvature favors the confinement of four $-1/2$ disclinations to the neck of the catenoid, the extrinsic curvature favors local orientation along the directions of zero or maximum curvature along the catenoid, which tends to repel defects from the neck.  As the relative strength of extrinsic to intrinsic effects is increased, this competition results in either a discontinuous or continuous expelling of the defects away from the neck, and in some cases off of the surface completely.  Anisotropic ordering on surfaces of complex curvature is common to many materials systems--from bicontinuous particle-stabilized emulsions, or ``bijels"~\cite{cates}, to biological membranes enriched by curvature-inducing proteins~\cite{fournier}--and we expect the frustration between extrinsic and intrinsic effects to play a generic role in shaping the structure and thermodynamics of these systems.

We begin by briefly overviewing the effect of intrinsic geometry on curved nematics.  We define an orthogonal coordinate system on the surface ${\bf u} = (u,v)$ and a corresponding orthonormal frame $\{\mathbf{e}_u,\mathbf{e}_v,\mathbf{N}\}$, where $\mathbf{N}$ is the unit surface normal. A two-dimensional nematic phase, therefore, has a director lying in the tangent plane such that $\mathbf{n}=\cos \theta({\bf u}) \mathbf{e}_u + \sin \theta({\bf u})  \mathbf{e}_v$.
For convenience, we choose a local isothermal, or conformal, coordinate system, $(u,v)$, defined by a metric of the form $ds^2=\Omega({\bf u}) (dv^2 + du^2)$.  In the presence of a configuration of disclinations, labeled by $n$ and having topological charges $s_n$ and positions $(v_n,u_n)$, one finds the explicit expression $\theta({\bf u}) = - \sum_n (s_n/2\pi) \mathrm{Im} \ln(u+i v-u_n-i v_n)$ and arrives at an energy written in terms of disclination position alone \cite{nelson_colloids, vitelli_turner}, 
\begin{eqnarray}
E_{in} &=& \frac{C}{ 2 }\sum_{m, n} s_n V_{int}({\bf u}_n  - {\bf u}_m ) s_m\\
& & + C \sum_n V_{G}({\bf u}_n) s_n \left(1-\frac{s_n}{4 \pi}\right),\nonumber
\end{eqnarray}
where $C$ is a constant modulus equal to the Frank constants in the one-constant approximation. The first term describes long-range interactions between disclinations, where $V_{int}({\bf u})$ is given by inverting the Laplace-Beltrami operator,  $\nabla^2 V_{int}({\bf u}) = \delta^2({\bf u})$.  The second term describes the interaction of the disclination with the Gaussian curvature, $G({\bf u})$, as encoded by a ``geometric potential,'' $V_G({\bf u})$, satisfying $\nabla^2 V_G({\bf u}) = G({\bf u})$~\cite{vitelli_turner}. This highlights the fundamental consequence of intrinsic geometry:  disclinations are attracted to regions of oppositely-signed Gaussian curvature \textit{via} long-range interactions.  

Thus, defects, through the intrinsic energy, ``screen'' Gaussian curvature, which would otherwise force lines of parallel orientation to converge or diverge. This is neatly demonstrated on the catenoid, a surface which concentrates negative Gaussian curvature within a narrow ``neck" adjoining two asymptotically planar surfaces.  Conveniently, the radial and azimuthal directions provide a natural isothermal coordinate system, $(v,u)$, respectively, with $\nabla^2 = [1/\Omega({\bf u})] (\partial_u^2 + \partial_v^2)$ and $G = - \nabla^2 (\ln \Omega)/2$. We immediately obtain $V_G ({\bf u}) = -\ln \Omega({\bf u})/2$.  For a catenoid with neck radius, $r$, we obtain the geometric potential explicitly, $V_G  ({\bf u })= -\ln \big[\cosh^2(v/r) \big]/2$.  The intrinsic energy thus favors negatively charged disclinations along the line of maximal curvature, $v=0$, and positively-charged defects are expelled.  The geometric potential is strongly confining for negative disclinations as $V_G \sim - |v|/r$ for $|v|\gg r$, indicating the divergent cost of expelling these disclinations to infinity.  Intuitively, the strong confinement of these defects can be understood from the uniform $\theta$ configurations shown in Fig. \ref{illustration} (b-c), which due to the absence of the requisite balance of negatively-charged defects in the neck effectively carry $+1$ disclination textures on each of the adjoining asymptotic planes.  Because $\int dA~G = -4 \pi$, a net defect charge of $-2$, or four $-1/2$ disclinations, are required to screen Gaussian curvature of the neck.

Extrinsic geometry further complicates this already complex picture. The extrinsic energy contributions to the free energy quantify the cost of out-of-plane gradients of ${\bf n}$ and are, therefore, sensitive to the surface's embedding, $\mathbf{X}({\bf u})$.  The surface curvature can be expressed as a tensor, $h_{i j} = \mathbf{N} \cdot \partial_i \partial_j \mathbf{X}$, where the indices can take on the values $v$ or $u$.  The eigenvectors of the tensor give the directions of maximal and minimal curvature while the corresponding eigenvalues give the principal curvatures. In the most general form, the coupling between $h_{i j}$ and the director can be expressed in terms of the three rotationally invariant quanitites: $C_{nn}=\mathbf{n}^i \mathbf{n}^j h_{i j}$, $C_{nt}=\mathbf{n}^i (\mathbf{N} \times \mathbf{n})^j h_{i j}$ and $C_{tt}=(\mathbf{N} \times \mathbf{n})^i(\mathbf{N} \times \mathbf{n})^j h_{i j}$~\cite{footnote}.  To second order in curvature, all terms allowed for achiral materials include~\cite{helfrich_prost} 
\begin{multline}\label{eq:extrinsic}
E_{ex} = \int dA\big\{ K_{nn} C_{nn}^2 + 2 K_{nt} C_{nt}^2 + K_{tt} C_{tt}^2 + \\
  + K'_{nn} C_{nn} + K'_{tt} C_{tt} \big\} .
\end{multline}
For a  microscopically symmetric interface such as a tilted-bilayer membrane, for which $\mathbf{N} \rightarrow - \mathbf{N}$ is a symmetry, terms linear in the curvature must vanish (i.e. $K'_{nn} = K'_{tt} =0$). For asymmetric interfaces, however, linear terms account for the physical distinction between  ``inside" and ``outside" \cite{stebe}.  Unlike the intrinsic energy, which is invariant under global rotations of the director around ${\bf N}$, the extrinsic coupling generates an unavoidable and geometrically-induced symmetry breaking field on anisotropic surfaces.  In terms of the angle field, 
\begin{equation}\label{Energy_reduced}
E_{ex}= \int dA \big\{ \gamma \cos [4 (\theta+\beta)]  + \gamma' \cos [2( \theta+\beta)]  \big\}  
\end{equation}
The strengths of the 4-fold and 2-fold symmetry-breaking fields are determined by the local anisotropy of bending on the surface,  $\gamma =  (K_{tt}+K_{nn}-2 K_{nt})(\kappa_1-\kappa_2)^2/8$  and and $\gamma' =  (K'_{nn}-K'_{tt})  (\kappa_1-\kappa_2)/2 + (K_{nn} - K_{tt})  (\kappa_1-\kappa_2)H$.  Here, the principal curvatures of the surface are $\kappa_1$ and $\kappa_2$, $H=(\kappa_1 + \kappa_2)/2$ is the mean curvature, and $\beta$ measures the angle between $\mathbf{e}_u$ and the direction of largest principal curvature.  For the case of the catenoid, note that both $H=0$ and $\beta=0$.

From this point of view, extrinsic energy leads to a generic preference to lock the director to the direction of local maximum or minimum curvature, a preference which is strongest in regions of high-curvature anisotropy where $\kappa_1-\kappa_2$ is large.  For example, the curvature distribution on the catenoid, $\kappa_1 = - \kappa_2 = r^{-1} {\rm sech}^2(v/r)$, leads to the extrinsic preference for uniform orientation most concentrated in the neck region.  Importantly, the uniformly-ordered ground states of the extrinsic energy on negatively-curved surfaces like the catenoid [Fig. \ref{illustration} (d-e)] correspond to the {\it maximal energy} configurations of the intrinsic energy.

To explore the defect configurations that best negotiate the compromise between intrinsic and extrinsic energies on the catenoid, we employ two methods.  First, we calculate the perturbative contributions from $E_{ex}$ in the limit cases of small $\gamma$ and $\gamma'$.  To first order, this amounts to evaluating Eq. (\ref{Energy_reduced}) at the intrinsic energy saddle-point with an additional global rotation of $\theta$ determined by intrinsic energy minimization for 4 disclinations distributed at intervals of $\pi/2$ around the neck and staggered at heights $v=\pm h$ away from the neck.   Second, we numerically study the ground state of a coarse-grained model of nematics on the catenoid.   Our approach is based on a generalization of the Lebwohl-Lasher model~\cite{lebwohl_lasher} for the case of a two-dimensional mesh of non-uniform geometry.

Starting with a catenoid discretized in a large number of small triangular patches,  a nematic director is assigned to the tangent plane of each patch.  The spin-spin interaction energy is given by
$H_{int} =  \sum_{i,j} \Delta A_i \Delta A_j \big[1 - ({\bf n}_i \cdot {\bf n}_j)^2 \big] V(r_{ij})$, where $V(r_{i j}) = \exp [-r_{i j}^2/(2 \sigma^2)]$ weights spin coupling between patches $i$ and $j$ by their spatial distance $r_{ij}$,  $\Delta A_i$ is the area of the $i^{th}$ patch and we choose $\sigma = 0.36 r$.  The area weighting has the important advantages that the coarse-grained model appropriately reduces to the continuum form of the intrinsic energy in the limit that $\Delta A_i \to 0$, and it minimizes the influence of nonuniform mesh geometry.  To this we add a finite-difference approximation of Eq. (\ref{Energy_reduced}) to account for extrinsic energies.   A Monte-Carlo simulated annealing scheme with the Metropolis-Hastings sampling method is used to determine the ground state 

\begin{figure}
\begin{center}
\includegraphics[width=3.in]{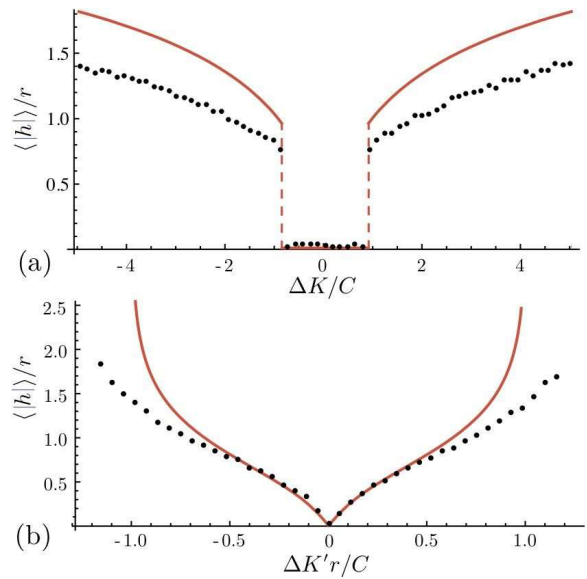} 
\caption{(color online) Mean vertical distance of the defects from the neck of the catenoid for a (a) symmetric and (b) asymmetric interface.  The solid curves are perturbation theory predictions, and the points are numerical simulation results. }
\label{height_plot}
\end{center}
\end{figure}

For symmetric interfaces, $\gamma'=0$, the relative strength of the extrinsic and intrinsic energy scales is characterized by $\Delta K /C$, where $\Delta K \equiv ( K_{nn} +K_{tt}-2 K_{nt})/8$.   In Fig. \ref{height_plot}(a), we plot the mean vertical distance measured from the neck at $v=0$ along the symmetry axis of the surface of the four -1/2 disclinations in the ground state.  Both the simulations and perturbation theory show that the defects are confined to $z=0$ by the geometric potential for small extrinsic couplings, $|\Delta K| < \Delta K_c \approx 0.8 C$.  At this critical coupling, the four disclinations are found to jump discontinuously to configurations outside of the high-curvature neck.  We denote this as a transition between the {\it confined} and {\it partially expelled} configurations, as the defects in the latter state, though expelled from the neck remain bound by the geometric potential and maintain a finite distance from the neck for any finite value of $\Delta K$ [Fig. \ref{illustration} (a)].

For an asymmetric interface with $\gamma=0$, we find dramatically different behavior.  Here, the relative strength of extrinsic to intrinsic energies depends on scale, characterized by the ratio $r \Delta K '/C$, where $ \Delta K ' =(K'_{nn}-K'_{tt})/2$.  As shown in Fig. \ref{height_plot}(b), instead of a discontinuous transition, the defects are pushed from the neck region for any finite value of $  \Delta K '$, with mean distance growing linearly as $r |\Delta K' |/C$ for weak extrinsic couplings.  Most significant, the four defects are expelled to $v \to \pm \infty$ as $ |\Delta K' |  \to \Delta K'_c \simeq C/r$ and are completely absent from the ground state of nematic order for stronger anisotropies.  In this case, we see a transition from a {\it partially expelled} to {\it fully expelled} state for the disclinations.

These two results for symmetric and asymmetric interfaces demonstrate the profound influence of extrinsic geometry on the ground state of nematic order. They also highlight, in different ways, the subtle relationship between disclinations and the anisotropic extrinsic curvature.
In the simplest view, the extrinsic anisotropy is localized to a finite region of size $r$ in the catenoid and, therefore, competes with the near-field isotropic texture that surrounds a defect core. A disclination in the neck costs an extrinsic energy of roughly $r^2 \Delta K (\Delta \kappa)^2 \approx \Delta K$ relative to the uniformly-oriented state. Relocating the disclination away from the neck adds an intrinsic energy penalty of order $s C$. Comparing, we expect defects to be expelled when $\Delta K \gg C$. However, $\int dA (\Delta\kappa)^2 = -\int dA G = 4 \pi$, so there is only a finite extrinsic energy to be gained;  disclinations can only be expelled partially.

\begin{figure}[t]
\begin{center}
\includegraphics[width=3.0in]{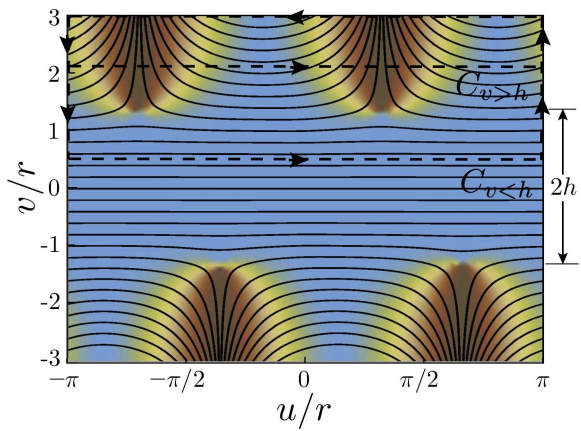} 
\caption{(color online) Conformal map of defects on a catenoid a distance $h$ from the center of the neck ($v=0$), shaded by extrinsic energy density with streamlines tangent to the nematic director. Integration contours $C_v<h$ and $C_v<h$ (dashed lines) have clockwise orientation.}
\label{conformal_map}
\end{center}
\end{figure}

This simple picture is not adequate on asymmetric interfaces.  Indeed, for nematics on asymmetric catenary interfaces, the extrinsic director coupling is far more homogeneous in its distribution.  This is straightforward to see in isothermal coordinates, where the area element grows exponentially with vertical height from the neck as $dA = du dv \cosh^2(v/r)$. This precisely balances the $|\Delta \kappa| = r^{-1} {\rm sech}^2(v/r)$ decay, indicating that extrinsic coupling is strong everywhere.  We can understand the far-field interactions between defects and extrinsic coupling by visualizing a partially-expelled nematic texture, with defects at $v=\pm h$, on the conformal map (the $v-u$ plane) (see Fig. \ref{conformal_map}). The intrinsic energy is $ \Delta K' r^{-1}\int du dv \cos [ 2 \theta ( {\bf u} ) ]$.  We consider contributions from constant-$v$ contours by examining the rotation of $\theta$ around closed contours, such as $C_{v>h}$ and $C_{v<h}$ shown in Fig. \ref{conformal_map} . In the map, the director rotates by $ + 2 \pi$ relative to the upper and lower boundary to maintain the uniform order in the asymptotic planes. The value of $\oint_c d {\bf \ell} \cdot {\nabla \theta} = 2 \pi - \pi n(v)$, where $n(v)$ is the number of disclinations in the region $v'>v$.   Since the contributions from the vertical portions of the contour cancel due to the $u \to u+ 2 \pi r$ symmetry, the director rotates by $+2 \pi$  along contours for $|v|\gtrsim h$, leading to no net extrinsic energy gain. For contours $|v| \lesssim h$, no net rotation of $\theta$ occurs, leading to a uniformly-oriented texture in the region spanning the disclinations and a coherent extrinsic energy gain of roughly $-  \Delta K' h$.   Comparing this to the intrinsic cost of pulling the defects from the neck, $ C h/r$, we find that the extrinsic effects are deconfining when $\Delta K' \gtrsim C/r$, and disclinations are expelled entirely from the surface.  Notice that the size scale of the surface becomes critically important for determining the ground state on asymmetric catenary surfaces.

We conclude by briefly noting that the competition between extrinsic and intrinsic effects is generic.  For example, we may consider the class radially-symmetric surfaces with the topology of the catenoid whose geometry far from the neck is described by $z(\rho\gg r) = r (r/\rho)^{\alpha}$, where $\rho$ is the radial distance from the axis and $z(\rho)$ is the vertical height from the neck.  Since $\kappa (\rho) \simeq \partial^2_\rho z \sim \rho^{-\alpha - 2}$ and the area element grows as $dA  \approx \rho d \phi d \rho$, it is straightforward to estimate the total extrinsic energy gain for expelling defects on asymmetric surfaces is, roughly, $\int dA \kappa \sim \int_r^R d \rho \rho^{-(\alpha+1)}$, where $R$ is the system size.  For $\alpha > 0$ this cost is finite and, when compared to the intrinsic energy cost to expel disclinations of $\ln (R/r)$, suggests that defects are always confined on the surface. On the other hand, the cost diverges when $R/r \rightarrow \infty$ for $\alpha < 0$ suggesting that defects are unstable for {\it any} finite extrinsic coupling.  Interestingly, the catenoid manifests the marginal case of $\alpha =0$, in which the partially-expelled and fully-expelled states are separated by a second-order critical point.  These simple arguments suggest that the ground-state thermodynamics of defects on curved interfaces is critically sensitive not only to the microscopic physics underlying the extrinsic and intrinsic couplings but also to subtle changes in surface geometry.

\begin{acknowledgments}
We thank R. Selinger for computational resources and fruitful discussions. This work was supported in part by an allocation of computing time from the Ohio Supercomputer Center. This work was supported as part of Polymer-Based Materials for Harvesting Solar Energy, an Energy Frontier Research Center funded by the U.S. Department of Energy, Office of Science, Office of Basic Energy Sciences under Award Number DE-SC0001087.
\end{acknowledgments}


\begin{thebibliography}{10}


\bibitem{vitelli_turner}
V. Vitelli and A. M. Turner, Phys. Rev. Lett {\bf 93}, 215301 (2004).

\bibitem{prost_lubensky}
T. C. Lubensky and J. Prost, J. Phys. II {\bf 2}, 371 (1992).


\bibitem{nelson_colloids}
D. R. Nelson, Nano Letters {\bf 2}, 1125 (2002). 


\bibitem{shin}
H. Shin, M. J. Bowick and X. Xing, Phys. Rev. Lett. {\bf 101}, 037802 (2008).

\bibitem{nelson_peliti}
D. R. Nelson and L. Peliti, J. Phys. {\bf 48}, 1085 (1987).

\bibitem{vitelli_nelson}
V. Vitelli and D. R. Nelson, Phys. Rev. E. {\bf 70}, 051105 (2004).

\bibitem{helfrich_prost}
W. Helfrich and J. Prost, Phys. Rev. A {\bf 38}, 3065 (1988).

\bibitem{terentjev}
P. Biscari and E. M. Terentjev, Phys. Rev. E {\bf 73}, 051706 (2006).

\bibitem{huber}
H. Jiang, G. Huber, R. A. Pelcovits and T. R. Powers, Phys. Rev. E {\bf 76}, 031908 (2007).


\bibitem{santangelo}
C. D. Santangelo, V. Vitelli, R. D. Kamien and D. R. Nelson, Phys. Rev. Lett. {\bf 99}, 017801 (2007).




\bibitem{footnote} These invariants can be constructed from the other two and invariants of the curvature tensor, $H$ and $G$.

\bibitem{stebe}
E. P. Lewandowski, J. A. Bernate, P. C. Searson and K. J. Stebe, Langmuir {\bf 24}, 9302 (2008).

\bibitem{lebwohl_lasher}
P. A. Lebwohl and G. Lasher, Phys. Rev. A {\bf 6}, 426 (1972).

\bibitem{cates}
E. Kim, K. Stratford, R. Adhikare and M. E. Cates, Langmuir {\bf 24}, 6549 (2008).

\bibitem{fournier} J.B. Fournier, Phys. Rev. Lett. {\bf 76}, 4436Ð4439 (1996).

\end{thebibliography}
\end{document}